\documentclass[prl,amssymb,showpacs]{revtex4}
\usepackage{graphicx}
\usepackage{dcolumn}
\usepackage{bm}
\usepackage{amsbsy}
\usepackage{amsmath}
\usepackage{amsfonts}

\begin{document}

\title{Electric field induced narrowing of exciton line width.}
\author{I. V. Ponomarev}
\email{ilya@bloch.nrl.navy.mil}
\author{L. I. Deych}
\author{A. A. Lisyansky}
 \affiliation{Department of Physics, Queens College of the City University of New
 York,\\
Flushing, NY 11367}
\date{\today }

\begin{abstract}
Considering effects of electric field on the low temperature
absorption line of quantum well excitons, we show that, for
moderate strength of the electric field, the main contribution to
the field dependence of the line-width results from field induced
modifications of inhomogeneous broadening of excitons. We find
that the strength of the random potential acting on quantum well
excitons due to alloy disorder and interface roughness can either
decrease or increase with field depending upon the thickness of
the well. This means that under certain conditions one can observe
counterintuitive narrowing of exciton spectral lines in electric
field.
\end{abstract}
\pacs{71.35.Cc,73.21.Fg, 78.67.De.}
 \maketitle



In the case of three-dimensional excitons, it is well known that
application of an electric field significantly reduces the exciton
life-time because of a finite probability of exciton tunnelling
through the field distorted potential barrier.
An obvious spectroscopic consequence of this effect is significant
broadening of exciton spectral lines at moderate electric fields.
However, excitons confined in a quantum well (QW) are much more
robust with respect to the electric field, ${\cal E}$, applied
perpendicular to the plane of a QW. As a result, QW excitons
demonstrate an appreciable field induced shift of their spectral
lines, while the spectral width does not change too much. The
physical origin of this quantum confined Stark effect
\cite{Miller85a} (QCSE) lies in the significantly increased
stability of QW excitons compared to the bulk case. Indeed, Stark
broadening  shows strong exponential dependence on field:
\begin{equation}\label{starkwidth}
\Gamma_{Stark}\sim E_0\exp\left(-2\hbar^2/3m|e|{\cal
E}\ell^3\right),
\end{equation}
where $E_0$ is a typical value of the resonance energy, $m$ is an
effective particle mass, and $\ell$ is a tunneling length. In QWs,
$\ell$ is determined by the confining potential of the well rather
than by the binding energy of excitons. Since the former is one or
two orders of magnitude greater than the latter, it is clear that
QW excitons in the perpendicular field can withstand much stronger
fields than their three dimensional counterparts. QCSE has
received a great deal of attention during the last two decades,
and was exploited in a number of electro-optic devices. However,
the issue of electric field induced changes in the exciton line
width, and their origin, still
remains largely unstudied (see 
Ref.~\onlinecite{Hong86a}).

In this paper we consider electric field induced modification of
exciton inhomogeneous broadening, which determines exciton line
width at low temperatures\cite{EfrosRaikh88,Herman91a,Runge02a}. We
show that modification of the random potential caused by a
reconstruction of electron-hole wave functions in the applied
electric field results in a {\em power law} field dependence of
exciton line width, which yields much stronger change in the line
width than the exponential dependence caused by Stark effect,
Eq.~(\ref{starkwidth}).

We find that for QWs whose thickness, $L$, is smaller than some
critical value, $L_{cr}$, an electric field actually reduces
fluctuations of this potential resulting in a counterintuitive
{\em narrowing} of the exciton line width with the electric field.
When $L>L_{cr}$, the sign of the electric field contribution to
the line width changes and exciton lines becomes broader with
field increase. The critical thicknesses, as well as $L$ and
${\cal E}$ dependencies of the exciton line width, are different
for compositional disorder and interface roughness mechanisms of
the inhomogeneous broadening. Therefore, experimental observation
of field induced changes in the low temperature exciton line width
can yield a unique method of characterization of QWs allowing
separation of these two contributions to the exciton's spectral broadening. 

Let us consider QW formed by a binary semiconductor, $AB$, as a
barrier material, and a ternary disordered alloy, $AB_{1-x}C_x$,
as a well. Throughout the paper we use effective atomic units: the
effective Bohr radius for length, $a_B=\hbar^2\epsilon/\mu^*e^2$,
$E_B=\mu^*e^4/\hbar^2\epsilon^2\equiv 2\textrm{ Ry}$ for energy,
and reduced electron-hole mass $\mu^*$ for masses,
$1/\mu^*=1/m_e^*+1/m_h^*$.
%
In the isotropic effective mass approximation, the Hamiltonian for
the exciton in QW with disorder is
\begin{equation}
H=H_0^e({\bf r}_e)+H_0^h({\bf r}_h)-|{\bf r}_e-{\bf
r}_h|^{-1}+U^e({\bf r}_e)+U^h({\bf r}_h),
\end{equation}
where $U^{e(h)}({\bf r}_{e(h)})$ are disorder induced potentials,
and $H_0^{e(h)}({\bf r}_{e(h)})$ are Hamiltonians for an
electron(hole) in QW:
\begin{equation}\label{H0_inQW}
H_0^{e(h)}({\bf r})=\frac{p_{e(h)}^2}{2m_{e(h)}}+
V_0^{e(h)}\theta\left(z^2-L^2/4\right)\mp
\frac{F^{e(h)}z}{2m_{e(h)}},
\end{equation}
where $F^{e(h)}=2m_{e(h)}|e|{\cal E}$, and $\theta(z)$ is the step
function. Inhomogeneous broadening of excitons in such a well is
determined by a combination of two types of disorder:
\begin{equation}
U^{e(h)}({\bf r}_{e(h)})=U^{e(h)}_{alloy}({\bf
r}_{e(h)})+U^{e(h)}_{int}({\bf r}_{e(h)}).
\end{equation}
Compositional disorder, $U^{e(h)}_{alloy}$, arising due to
concentration fluctuations in a ternary component, which produce
local band gap fluctuations
\cite{EfrosBaranov78,Zimmer92a,EfrWetWor95}, and interface
roughness, $U^{(e,h)}_{int}$, caused by the formation of monolayer
islands on the QW interfaces that results in local changes in the
QW thickness
\cite{Weisbuch81a,Singh84a,Srinivas92a,Zimmermann97a}. 

Usually, the exciton binding energy in QW is much larger than
disorder-induced local energy fluctuations. Therefore excitons are
expected to move through QW as a whole entity. It is formalized in
a representation of the total wave function of the electron-hole
pair in the form of a product,
\begin{equation}
\Psi({\bf r_e,r_h})=\Phi({\bf R})\psi({\bm
\rho})\chi_{e}(z_{e})\chi_{h}(z_{h}),
\end{equation}
where ${\bf r}_{h,e}=({\bm\rho}_{h,e};z_{h,e})$,
${\bm{\rho}}={\bm\rho}_e-{\bm\rho}_{h}$,
$\textbf{R}=(m_e{\bm\rho}_{e}+m_h{\bm\rho}_{h})/M$, $\Phi({\bf
R})$ is a wave function for center-of-mass lateral motion,
$\psi({\bm \rho})$ is an exciton relative lateral motion wave
function, and $\chi_{e,h}(z_{e,h})$ are one-dimensional  electron
and  hole QW ground state wave functions. Functions $\psi$ and
$\chi$ are solutions of the corresponding Schr\"{o}dinger equation
for a perfect QW without disorder, while the Schr\"{o}dinger
equation for the center-of-mass motion includes effective random
potentials, $V_{eff}({\bf R})=V_{int}({\bf R})+V_{alloy}({\bf
R})$, obtained from the averaging of the original random
potentials $U^e({\bf r_e}),\ U^h({\bf r_h})$ over ${\bm \rho}$ and
$z$ coordinates. (In these calculations we do not take into
account disorder-induced renormalization of functions $\psi$,
$\chi_{e,h}$ and the corresponding energies \cite{EfrosRaikh88},
which results in effective decreasing of Bohr's radius $\lambda$.
This effect does not change qualitative conclusions of our work
[see Eqs.~(\ref{alloydis_gen}) and (\ref{interface_gen})].)
%

Our primary goal is to calculate the variance of the effective
random exciton potential defined as
\begin{equation}
W=\sqrt{\langle V_{eff}({\bf R})^2\rangle}.
\end{equation}
This
parameter determines a number of experimentally observable
quantities such as the exciton radiative life-time and the
absorption line width \cite{EfrosRaikh88,Herman91a,Runge02a}.
  The radiative life-time can be extracted from the exciton
absorption line-shape \cite{EfrWetWor95}, whose calculation in the
dipole approximation is equivalent to the estimation of the
optical density function: $A(\varepsilon) =
\left\langle\sum_i\left|\int
d^2\,R\Phi_i(\textbf{R})\right|^2\delta(\varepsilon-\varepsilon_i)\right\rangle.
$
The interpolation procedure \cite{EfrosRaikh88,EfrWetWor95}
gives an asymmetric line shape towards high frequencies. In many
cases however, for estimation of the line-width it is reasonable
\cite{Herman91a,Runge02a} to consider that the underlying
disorders are described by the Gaussian random processes. Then the
shape of the exciton line is also Gaussian,
and the corresponding full-width-at-half-maximum is given by
$\Delta=\sqrt{8\ln(2)}W$.

Since contributions from the alloy and the interface disorders can
be considered statistically independent of each other,
$W_{tot}^2=W_{alloy}^2+W_{int}^2$.
Estimations show that usually both disorders
yield comparable contributions to the total width. This imposes an
additional difficulty for experimental identification of the
interface quality in QWs from optical spectra, since absolute values
of each contribution are usually unknown.

 The effective random potentials acting on the
exciton's center-of-mass, for each type of disorder, is presented as
a sum of two terms $V({\bf R})=V_h({\bf R})+V_e({\bf R})$,
representing hole and electron contributions,
\begin{equation}\label{randpotR}
V_{h,e}=\int U_{h,e}({\bf R}\pm
m_{e,h}{\bm\rho}/M;z)\psi^2({\bm{\rho}})\chi^2_{h,e}(z)\; d^2\rho
dz.
\end{equation}
Correspondingly, each variance will have three terms
 $W^2=\langle V_{h}^2 +2V_{h}V_{e}+V_{e}^2\rangle.$
For a QW with a heavy hole and light electron $(m_h\gg m_e)$ as in
$In_xGa_{1-x}As/GaAs$ heterostructures, the main contribution
stems from the hole-hole part due to the enhancement factor
$(M/m_e)$ which appears after averaging over lateral coordinates
${\bm\rho}$ in Eq.~(\ref{randpotR}). Considering only this case,
we neglect terms containing a $V_e$ factor.  Then the microscopic
potential representing the alloy can be presented as (hereafter
the subscript ``$h$" is omitted) \cite{EfrWetWor95}
\begin{equation}\label{alloydispot}
U_{alloy}({\bf r})=\alpha\xi({\bf
r})\theta\left(L^2/4-z^2\right)/N,
\end{equation}
where $\theta(z)$ is a step-function, $N$ is the concentration of
lattice sites ($N=4/a_{lat}^3$ for zincblende materials, $a_{lat}$
is a lattice constant), $\xi({\bf r})$ is the random fluctuation
of the local concentration of atoms in the alloy from the average
value $xN$, and $\alpha=dE_v/dx$ characterizes the rate of shift
of the valence bands with composition $x$. The interface roughness
potential can be presented in the following form
\cite{Zimmermann97a,pon04a}
\begin{equation}\label{Uinterf}
U_{int}(\textbf{r})=
V_0\left[\eta_1({\bm\rho})\delta(z+L/2)-\eta_2({\bm\rho})\delta(z-L/2)
\right],
\end{equation}
where $\delta(z)$ is a $\delta$-function, $V_{0}$ is a hole
off-set band energy. Random functions $\eta_{1,2}({\bm\rho})$ with
zero mean characterize a deviation of the $i$th interface from its
average position.

The statistical properties of alloy and interfacial roughness are
characterized by the correlators
\cite{EfrosBaranov78,Zimmermann97a,Meyerovich98,Meyerovich99,pon04a}:
\begin{eqnarray}
\langle \xi({\bm{r}}_1) \xi({\bm{r}}_2)\rangle &=&
x(1-x)N\delta(|{\bm{r}}_1-{\bm{r}}_2|),\label{alloycor}\\
\langle \eta_i({\bm{\rho}}_1) \eta_j({\bm{\rho}}_2)\rangle &=&
h^2f_{ij}\zeta(|{\bm{\rho}}_1-{\bm{\rho}}_2|),\label{intcor}
\end{eqnarray}
where $h$ is an average height of interface inhomogeneity, and
$\langle\cdots \rangle$ denotes an ensemble average. For the
interface height-height correlator we assume that the dependence of
both diagonal and non-diagonal correlations on the lateral
coordinates ${\bm \rho}$ is described by the same function
$\zeta({\bm \rho})$. The diagonal elements $f_{ii}$ are different if
two interfaces are grown under different conditions, which happens
naturally for $GaAs$ based structures. (Growth of a ternary alloy on
$GaAs$ occurs differently from growth of $GaAs$ on the alloy;
besides using techniques of growth interruption one can
significantly modify statistical properties of the grown
interfaces.) The non-diagonal element $f_{12}(L/\sigma_{\|})$
introduces correlations between different interfaces. The respective
quantity, which can be called the \emph{cross-} or
\emph{vertical-correlation function}
\cite{Meyerovich98,Meyerovich99,Meyerovich02,pon04a}, is a function
of the average width of the well and is characterized by the
vertical correlation length $\sigma_{\|}$. The presence of these
correlations suppresses the interface disorder contribution into
inhomogeneous broadening, especially, for narrow QWs
\cite{pon04a,pon05a}.

\begin{figure}[h]
\includegraphics{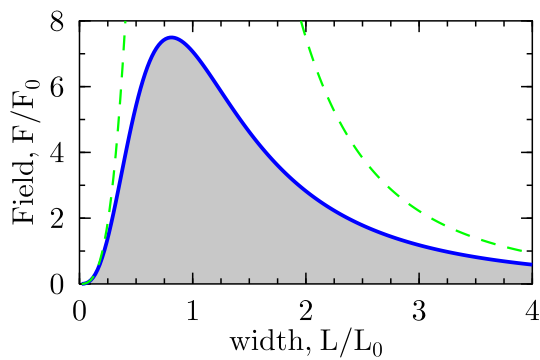}
 \caption{(Color online) Phase diagram $F$-$L$ of dominant contributions to the
exciton line-width: the gray shaded region - disorder induced
mechanisms, the outside region - Stark broadening. Dashed lines are
the shallow well and the infinite well approximations. The  QW width
is measured in units $L_0=\pi/u$, which determines the number of
levels in the QW, and $F_0$ defines the natural scale for electric
field (see text). } \label{fig1}
\end{figure}

In order to calculate the effective potential, one needs to know the
exciton wave functions $\psi$ and $\chi$ for an ideal QW in the
perpendicular uniform electric field. They can be found with the
help of the variational method. It is a well-known fact
\cite{Miller85a,deych04a} that lateral relative motion is very
weakly affected by perpendicular electric field. The corresponding
trial function can be chosen in a form of the hydrogen $1S$-like
orbital, $\sqrt{2/\pi\lambda^2}\exp(-r/\lambda)$, with the
quasi-two-dimensional Bohr's radius $\lambda$ as a variational
parameter. In principle, the one-dimensional single-particle
function $\chi(z)$ for a hole in QW, which satisfies the
Schr\"{o}dinger equation
\begin{equation}
\left[ p_z^2/2m +
V_0\theta\left(z^2-L^2/4\right)+Fz/2m\right]\chi=E\chi,
\end{equation}
can be found exactly in terms of the Airy
functions. This solution corresponds to a quasi-stationary hole
state, and describes a possibility for the hole to tunnel out of the
well. It diverges at infinity, and is not suitable, therefore, for
calculation of the effective potential, Eq.~(\ref{randpotR}). In
this paper, however, we are interested in the range of parameters
where Stark broadening is small.
This restricts
our consideration to a certain region  on the $F$-$L$ plane (grey
shaded area in Fig.~\ref{fig1}), where the exponent of
Eq.~(\ref{starkwidth}) is smaller than unity.
It is worth to note, that these fields can reach very high values
since the relevant energy scale is the confining QW potential
rather than the Coulomb interaction between the hole and the
electron. For example, for In$_{0.18}$Ga$_{0.82}$As/GaAs QWs the
maximum in Fig.~1 corresponds to the field $8\times10^{5}$ V/cm.

Previous studies of QCSE showed 
\cite{Nojima88a} that within this range of parameters, an
approximation of $\chi(z)$ by a real function, which can be found
with the help of variational method, gives a very good description
of both the energy and the wave function of a hole in the presence
of a perpendicular electric field. Reasonable results can be
obtained even for the simplest one-parameter variational function of
the following form \cite{deych04a}:
\begin{equation}\label{chiQW}
\chi(z;\beta)=B(\beta,k)\exp(-\beta z)\chi_0(z),
\end{equation}
where the wave function, $\chi_0(z)$, represent the hole ground
state in QW without the electric field:
\begin{equation}
\chi_0(z)=\left\{
\begin{array}
[c]{cc}%
\cos(kz), & z\leq |L/2|\\
\exp\left[-\kappa (|z|-L/2)\right], & z\geq |L/2|,
\end{array}
\right. \label{chiQW0}
\end{equation}
$B$ is the normalization constant:
\begin{equation}
B(\beta,k)=\sqrt{\frac{2\beta(\kappa^2-\beta^2)(k^2+\beta^2)}{k^2\left[
2\kappa\beta\cosh(\beta L)+(\kappa^2+\beta^2)\sinh(\beta
L)\right]}},
\end{equation}
and we introduced the following notations $k=\sqrt{2mE}$,
$u=\sqrt{2mV_0}$, and $\kappa=\sqrt{u^2-k^2}$. The wave number $k$
is given by a root of the transcendental equation,
$kL/\pi=1-(2/\pi)\arcsin(k/u)$. Eq.~(\ref{chiQW0}) for $\chi_0(z)$
guarantees a continuity of the wave function and its derivative at
interfaces $z=\pm L/2$.
%
Parameter $u$ defines a natural length scale for the QW width,
$L_0=\pi/u$. It counts the number of levels in QW: $[L/L_0]=n+1$.

In general, the solution for the variational parameter
$\beta(F,L,u)$ can be found only numerically. One can show,
however, that for a moderate field, $\beta$ is proportional to the
electric field, $\beta=C(L)F$, where
\begin{equation}\label{CL}
C(L)=\frac{1}{\kappa^2}-\frac{1}{k^2}
     +\frac{L}{\kappa}\frac{1+L\kappa+L^2\kappa^2/6}{2+L\kappa}.
\end{equation}
The constant $C(L)$ introduces the natural scale for electric field
units $F_0=C(L)^{-3/2}\equiv\ell^{-3}$, where $\ell$ defines an
average extension of wave function in QW. Analytical expression for
$C(L)$ can be obtained in two important limits: a very wide well,
which can be approximated by an effective infinite QW
\cite{Nojima88a} with $k\approx \pi/L$, $\kappa\approx u$, and a
very narrow shallow QW \cite{deych04a}, which can be described by a
model of a $\delta$-functional QW potential with $k\approx u$,
$\kappa\approx mV_0L$:
\begin{equation}
C_{\infty}=\frac{L^2}{2}\frac{\pi^2-6}{6\pi^2},\
C_{\delta}=\frac{1}{2(mV_0L)^2}.
\end{equation}
Using correlators, Eqs.~(\ref{alloycor}) and (\ref{intcor}), and
the wave functions $\psi(\rho)$ and $\chi(z;\beta)$ we obtain the
following expressions for alloy and interface roughness variances:
\begin{equation}\label{alloydis_gen}
 W^2_{alloy}= \frac{a_{lat}^3x(1-x)}{8\pi\lambda^2}
  \frac{\alpha_h^2M^2}{m_e^2}\int_{-L/2}^{L/2}\chi(z;\beta)^4\,dz,
\end{equation}
\begin{equation}\label{interface_gen}
 W^2_{int}=V_0^2h^2G(y)\left[
 f_{11}\chi_L^4+f_{22}\chi_R^4-2f_{12}\chi_L^2\chi_R^2\right],
\end{equation}
where $\chi_{L,R}\equiv\chi(\mp L/2;\beta)$. The function $G(y)$,
defined as \cite{pon04a}
\begin{equation}
G(y) = \int\;d^2\rho d^2\rho'\psi^2(\rho)\psi^2(\rho')
\zeta\left(|{\bm\rho'}-{\bm\rho}|m_e/M\right),\label{T3form}
\end{equation}
with $y=\sqrt{2}\sigma_{\perp}M/(\lambda m_e)$, depends on the
lateral correlations of interface roughnesses, characterized by
the in-plane
correlation radius $\sigma_{\perp}$.

Since, for small $F$, parameter $\beta$ is proportional to the
field, both Eqs.~(\ref{alloydis_gen}) and (\ref{interface_gen})
can be expanded in terms of $F$:
\begin{eqnarray}
 W^2_{all}(F)&\approx& \Omega_{all}
 \left[\gamma_0^{(all)}+\gamma^{(all)}_2F^2\right],\label{alloyF}\\
W^2_{int}(F)& \approx & \Omega_{int} \left[\gamma_0^{(int)}+
\gamma_1^{(int)}F+\gamma_2^{(int)}F^2\right],\label{interF}
\end{eqnarray}
where
\begin{eqnarray}
\Omega_{all}&=&\left[\kappa^2 a_{lat}^3x(1-x)\alpha_h^2M^2
\right]/\left[(\kappa L+2)^2 2\pi\lambda^2 m_e^2\right],\\
\Omega_{int}&=&4\kappa^2k^4V_0^2h^2G(y)/[(\kappa L+2)^2u^4],\\
\gamma_0^{(int)}&=&f_{11}+f_{22}-2f_{12}(L),
\end{eqnarray}
and all other $\gamma_i$ are also monotonic functions of the QW
width.
 Eqs.~(\ref{alloyF}) and (\ref{interF}), which present the main
results of the paper, show that in the range of parameters where the
Stark width is exponentially small, there exists a strong
\emph{power law} field dependence of inhomogeneous exciton
broadening caused by the field induced changes in the variance of
the effective exciton potential.

The first remarkable feature of Eqs.~(\ref{alloyF}) and
(\ref{interF}) that we would like to point out is the presence of
thw linear-in-field term (see Fig.~2) in the interface roughness
\begin{equation}
\gamma_1^{(int)}(L)=2LC(L)\left(f_{11}-f_{22}\right).
\end{equation}
One can see that this term results from asymmetry between two
interfaces of the well, which manifests itself through different
roughnesses, $f_{11}\neq f_{22}$. In $GaAs$ based
heterostructures, this asymmetry appears naturally because of the
polar nature of $GaAs$, but it can also be engineered by preparing
different interfaces under different growth conditions. The
presence of the linear term gives rise to an interesting effect:
one can switch between field induced narrowing or broadening of
the exciton line by simply changing the polarity of the applied
field. This effect has a simple physical interpretation: if QW
holes are pushed by the field toward a less disordered interface,
the exciton line narrows, but it broadens in the opposite
situation.
\begin{figure}[h]
\includegraphics{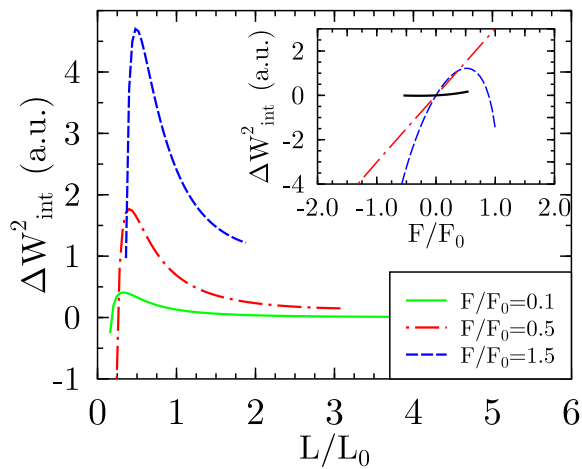}
 \caption{(Color online) The field dependent part of the variance,
 $\Delta W_{int}^2=\Omega\left[\gamma_1F +\gamma_2 F^2\right]$, as a function of the QW width for
the interface roughness contribution with different corrugation at
interfaces ($f_{11}=4, f_{22}=1$) for three values of the
 electric field: $F/F_0=0.1, 0.5, 1.5$.
  Curves are drawn only for the gray shaded area of Fig.~1, where
  disorder contributions to the exciton line width dominate.
  Inset: $\Delta W_{int}^2$ as a function of electric field for
  three QW widths: $L/L_0=0.3$ (dashed), $L/L_0=0.55$
  (dotted-dashed), and $L/L_0=3$ (solid). Note that $L/L_0=0.55$
  corresponds to the case, when the second order in field term
  disappears (see text).
 } \label{fig2}
\end{figure}

\begin{figure}[h]
\includegraphics{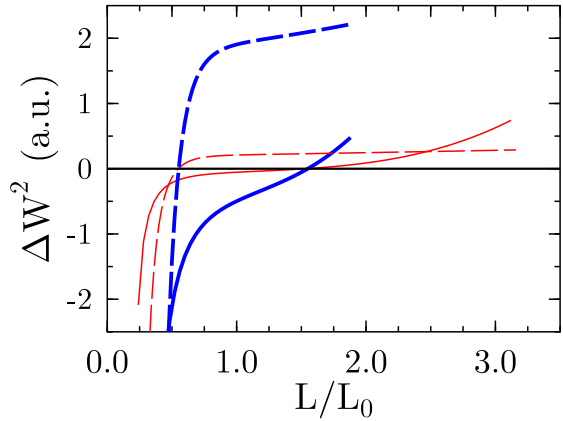}
 \caption{(Color online) The field dependent part of the variance,
 $\Delta W^2=\Omega \gamma_2 F^2$, as a function of the QW width for alloy disorder (solid curves)
 and identical ($f_{11}=f_{22}$) interface roughness (dashed curves) for
 two values of the electric field: $F/F_0=0.5$ (thinner curves) and $F/F_0=1.5$
(thicker curves).
  Curves are drawn only for the gray shaded area of Fig.~1.
 } \label{fig3}
\end{figure}

Quadratic in the field terms in Eqs.~(\ref{alloyF}) and
(\ref{interF}) also possess nontrivial properties (see
Fig.~\ref{fig3}). In the limit of shallow $\delta$-functional QWs,
factors $\gamma_2(L)$ can be presented as
 \begin{eqnarray}
\gamma^{(all)}_2&=&-2C_{\delta}^2\left[(mV_0L)^{-2}-L^2/3\right],\\
\gamma^{(int)}_2&=&-2C_{\delta}^2\left[\frac{\gamma_0^{(int)}}{(mV_0L)^{2}}
-2L^2\left(\gamma_0^{(int)}+4f_{12}\right)\right].
\end{eqnarray}
At small widths, $L<L_{cr}$, the first term in square brackets in
both equations dominates making respective contributions to the
line width negative. In the opposite limit of an effective
infinite wall, these factors are positive:
\begin{eqnarray}
\gamma^{(all)}_2&=&C_{\infty}^2L^2\left[1/3-3/\pi^2\right]\\
\gamma^{(int)}_2&=&C_{\infty}^2\left[\gamma^{(int)}_0
 \left(5/3+2/\pi^2 \right)+ 4f_{12}\left(1+2/\pi^2 \right)\right].
\end{eqnarray}

Different signs corresponding to the opposite limits mean that at
some particular QW widths, factors $\gamma_2(L_{cr})$ vanish.
Numerical analysis shows (see Fig.~\ref{fig3}) that these
``critical" widths are different for alloy disorder and interface
roughness contributions. For parameters used in constructing
Fig.~\ref{fig1}, $L_{cr}$ for alloy disorder corresponds to
$L/L_0\approx 1.5$, and for interface disorder to $L/L_0\approx
0.55$. This means that we can effectively turn off the quadratic
contribution from one of two sources of the inhomogeneous
broadening by growing QW with a width close to the respective
critical value, $L_{cr}$. In this case, all the field induced
changes in the exciton broadening will be caused mostly by the
other broadening mechanism. In principle, this can allow for
unambiguous discrimination between alloy and interface disorder
contributions to the exciton line width. For example, growing QW
with a size $L/L_0\approx 1.5$ and measuring field-induced changes
in the exciton line-width, we can guarantee that that these
changes originate from the interface roughness mechanism only,
since contribution from the alloy disorder mechanism vanishes up
to the third order in field terms (see the dotted-dashed line in
Fig.~4).
\begin{figure}[h]
\includegraphics{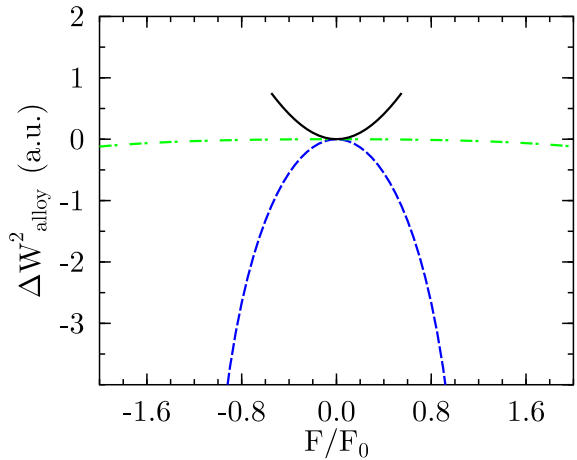}
 \caption{(Color online)
The electric field induced change of the exciton variance $\Delta
W_{alloy}^2$ for the alloy disorder contribution for three
different QW widths: $L/L_0=3$ (solid line), $L/L_0=1.5$
(dotted-dashed line) and $L/L_0=0.3$ (dashed line).  Curves are
drawn only for the gray shaded area in Fig.~1. Contribution is
negative for small thicknesses, and positive at larger
thicknesses. The critical value of QW width is determined as a
moment when the contribution changes its sign:
$L_{cr}^{alloy}\approx 1.5L_0$.
 } \label{fig4}
\end{figure}
Different signs of $\gamma_2$ for shallow and deep QWs can be
explained by a competition of two processes. On the one hand, the
electric field pushes part of  the wave function outside of the
well and away from the influences of the disorders, promoting
narrowing of the exciton line. On the other hand, the field
changes a shape of the wave function, pushing it towards an
interface and slightly squeezing. The latter results in greater
localization of the wave function and hence, broadens the exciton
line. It is clear that the first process dominates for shallow
QWs, while the second one prevails for  QWs with larger widths.

To estimate quantitatively the critical QW widths we chose the
example of In$_{0.18}$Ga$_{0.82}$As/GaAs. For this QW the material
parameters are: $m_e^{*}=0.052m_0$, $m_{h}^{*}=0.31m_0$,
$\mu_X^{*}=0.045m_0$, $U_h=79\,$meV, $E_B=6.7\,$meV, $a_B=16\,$nm.
 We obtained $L_{cr}^{int}=1.5\,$nm and $L_{cr}^{alloy}=4.2\,$nm  for critical interface and
alloy disorder lengths respectively. Quantitative values for QWs
made from different materials can be readily recalculated using
dimensionless values of field and length in Figs. 1-4 and the
following formulas for electric field and QW lengths in standard
units:
\begin{eqnarray}
L[\textrm{nm}]&=&a_B[\textrm{nm}]\frac{L}{L_0}\sqrt{\frac{\pi^2\mu^*E_B}{2m_h^*V_0^h}}\nonumber\\
F[\textrm{V/cm}]&=&10^6F_0\frac{E_B[\textrm{meV}]}{a_B[\mbox{nm}]}
\frac{\mu^*}{2m_h^*}\frac{F}{F_0},
\end{eqnarray}
where $F_0$ can be extracted from Eq.~(\ref{CL}).


In conclusion, we considered effects of an electric field on the
inhomogeneous line width of QW excitons. It is shown that
interface roughness can result in linear with respect to the field
contribution to the exciton line width. This gives rise to the
effect of switching between narrowing and broadening of the
exciton line by changing field polarity. Quadratic contributions
to the field dependence of the line width from both alloy and
interface disorders are negative for shallow QWs, but change signs
with the increase of the QW depth. These effects reveal a rich
physics of inhomogeneously broadened excitons in an electric
field, and can be used in applications for narrowing exciton
spectral lines and experimental study of the roles of different
mechanisms of inhomogeneous broadening of excitons. We would like
to stress that even though we considered a
 simplified model of QW, the incorporation  of such effects  as valence-band mixing, non-parabolicity of
the conduction band, dielectric constant and effective mass
mismatches  would not change the qualitative results of the paper,
which are related to the presence of two competing mechanisms
affecting the exciton effective potential rather than to any
particular model of the electron (hole) band structure or the
confinement potential.
\begin{acknowledgments}
 We are grateful to S. Schwarz for reading and commenting on the
manuscript. We thank A. Efros for helpful discussions. The work is
supported by AFOSR grant F49620-02-1-0305 and PSC-CUNY grants.
\end{acknowledgments}

\end{document}